# Assessing contribution of treatment phases through tipping point analyses using rank preserving structural failure time models

Sudipta Bhattacharya and Jyotirmoy Dey

**Abstract:** In clinical trials, an experimental treatment is sometimes added on to a standard of care or control therapy in multiple treatment phases (e.g., concomitant and maintenance phases) to improve patient outcomes. When the new regimen provides meaningful benefit over the control therapy in such cases, it proves difficult to separately assess the contribution of each phase to the overall effect observed. This article provides an approach for assessing the importance of a specific treatment phase in such a situation through tipping point analyses of a time-to-event endpoint using rank-preserving-structural-failure-time (RPSFT) modeling. A tipping-point analysis is commonly used in situations where it is suspected that a statistically significant difference between treatment arms could be a result of missing or unobserved data instead of a real treatment effect. Rank-preserving-structural-failure-time modeling is an approach for causal inference that is typically used to adjust for treatment switching in clinical trials with time to event endpoints. The methodology proposed in this article is an amalgamation of these two ideas to investigate the contribution of a treatment phase of interest to the effect of a regimen comprising multiple treatment phases. We provide two different variants of the method corresponding to two different effects of interest. We provide two different tipping point thresholds depending on inferential goals. The proposed approaches are motivated and illustrated with data from a recently concluded, real-life phase 3 cancer clinical trial. We then conclude with several considerations and recommendations.

# 1. Introduction

Adding a new experimental drug to a standard of care (SOC) treatment regimen to improve efficacy in a disease is common practice in medical research. A randomized clinical trial (RCT) is usually needed to compare the new regimen to the SOC regimen to evaluate its comparative efficacy and safety. When the experimental drug is utilized in multiple ways within the same regimen, such as in multiple phases, it raises questions about the necessity of each of these phases even if the overall regimen is shown to outperform the SOC. In such cases, the effect of the component phases can however be confounded in a way that the contribution of each phase to the overall efficacy is difficult to discern. For example, Stupp et al.[1] reported superior overall survival results compared to SOC radiation therapy (RT) in subjects with newly diagnosed glioblastoma multiforme of a new treatment regimen which administered a new drug temozolomide (TMZ) concomitantly with RT at a lower dose for 6 weeks followed by six adjuvant cycles of TMZ alone at a higher dose. The contribution of each of the two treatment phases with TMZ to the overall efficacy is however not separable or fully understood.

The motivating and illustrative example for our proposal comes from BROCADE3, a recently conducted phase 3 clinical trial of veliparib, a new inhibitor of the enzyme poly ADP ribose polymerase (PARP), in subjects with breast cancer that tested negative for human epidermal growth factor receptor 2 (HER2 negative) genes but had mutated tumor suppressor genes BRCA1 and BRCA2 (BRCA-mutated). Subjects on study received treatment in a (chemo-) combination phase comprising standard of care (SOC) chemotherapy (CT) along with veliparib (or placebo) with the option to continue veliparib at a higher dose (or placebo) in a maintenance monotherapy phase when SOC CT is discontinued. For reasons explained in the next section, it was important in this context to understand the contribution of the combination phase to the full regimen.

In general, when a new therapy A is added to standard of care therapy C to form a new treatment regimen for a given disease, the most reliable way to evaluate its efficacy is to conduct a randomized controlled clinical trial of A+C vs C. However, when two new therapies A and B are added on to the control C, it is usually not adequate to simply show that A+B+C is more efficacious than C without supportive evidence that both A and B are important contributors to the observed improvement. Also, in general, it is not possible to isolate the contribution of either A or B from a two-arm RCT of A+B+C vs C without making additional assumptions and/or using external data. If there is skepticism about the contribution of A to the efficacy of A+B+C, one way to assess it is to compare A+B+C vs. B+C in an RCT. To assess whether B by itself is sufficiently efficacious (i.e., A is not essential), one must compare B+C vs. C. (Figure 1)

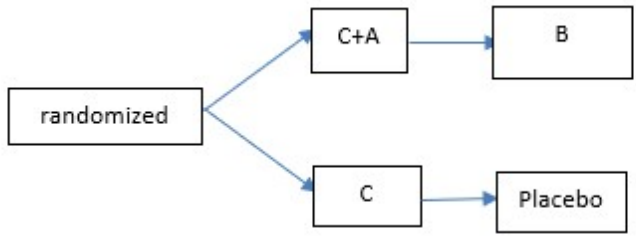

**Figure 1A.**     **Original treatment scheme for the study**

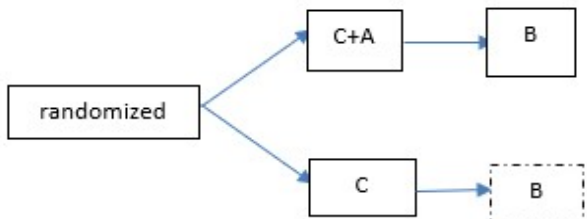

**Figure 1B.** Counterfactual treatment scheme with elicitation made over the placebo arm

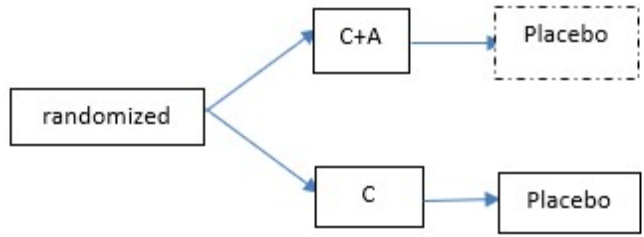

**Figure 1C.** Counterfactual treatment scheme with elicitation made over active treatment arm

**Figure 1.** Design of the treatment scheme for the study and after counterfactual elicitation

Evaluating either of these comparative effects may be important in the context of considering A+B+C as a new treatment for patients. For example, when A+B+C is less safe than C alone, it may be worthwhile reducing the treatment burden on patients (and possibly improving the risk-benefit profile of the new therapy) by eliminating A (or B) if its contribution to the efficacy of the full regimen is not substantial. It is also important to note that A and B does not need to be different drugs for the statements above to hold. They hold even if A and B are two different schedules, strengths, or formulations of the same drug. When A and B are administered in separate treatment phases, their effects are, to some extent, temporally separated and one may try to use it to gain additional insights.

In this article we propose a novel tipping point analysis (TPA) using rank preserving structural failure time (RPSFT) modeling to evaluate the influence of a specific temporally separated component phase within a treatment regimen. The method is described by developing the framework and applying it to assess the contribution of the combination phase in the BROCADE3 example. In subsequent sections, we will provide further background information about the BROCADE3 study and its key published findings (Section 2), describe the mathematical framework for our proposed TPA methodology in detail (Section 3), apply the method to the data from BROCADE3, explore variants of the method depending on the effect of interest (Section 4), and discuss pros and cons of using the method and provide recommendations on the best way to use it (Section 5).

## 2. Motivating Example: BROCADE3

BROCADE3 is a phase 3 randomized, double-blinded, placebo-controlled study of veliparib in subjects with BRCA-mutated HER2-negative breast cancer (Diéras et al.[2]). Subjects were randomized in a 2:1 fashion to receive either veliparib or placebo added on to SOC chemotherapy (CT) with carboplatin and paclitaxel. (Figure 2) While treatment with all three drugs (carboplatin, paclitaxel and veliparib/placebo) could continue until disease progression or unacceptable toxicity, study subjects and their treating physicians had the option to discontinue any of these drugs at any time. When both SOC chemotherapy drugs were discontinued without disease progression or death, subjects had the option to receive monotherapy maintenance with veliparib/placebo (blinded) at a higher dose.

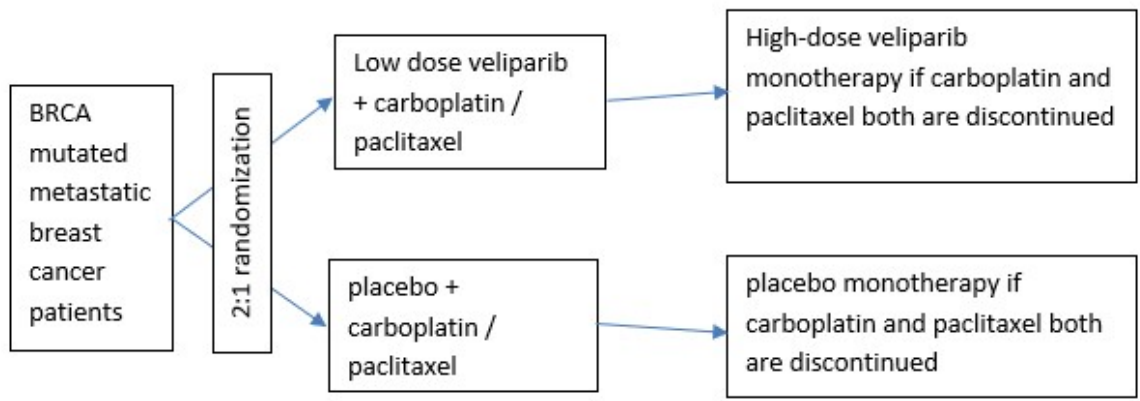

**Figure 2.    BROCADE-3 Study Design**

Primary findings of this study showed statistically significant improvement in progression-free survival (PFS) (Diéras et al.[2]) and late separation of Kaplan-Meier (KM) curves for the two treatment arms (Table 1, Figure 3). Subjects on study received 7.5 months of SOC CT on average (median 6.3 months) and ultimately about 36.9% of subjects received veliparib monotherapy maintenance at a higher dose. A total of 349 subjects had experienced at least one PFS event by the time of the primary analysis and hence, the minimum detectable difference (MDD) at the 2-sided 5% level of significance was approximately a hazard ratio of 0.8.

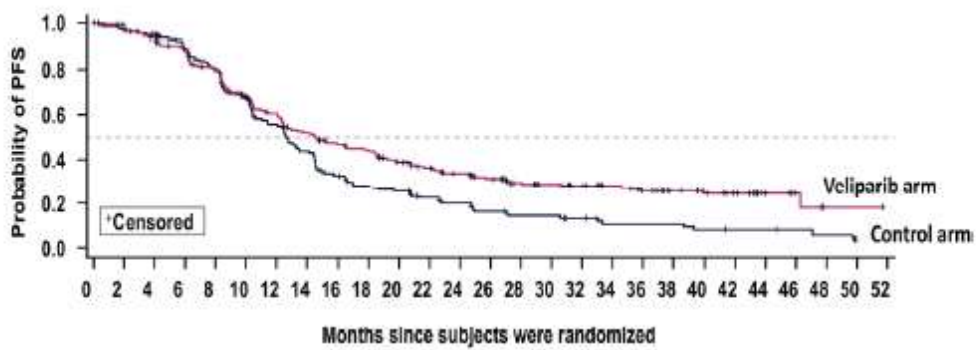

**Figure 3 Kaplan-Meier survival probability plot for progression free survival from BROCADE 3 study**

**Table 1:    Primary analysis result**

|  | Experimental Arm (n=337) | Control Arm (n=172) |
|---|---|---|
| PFS events | 217 | 132 |
| Censored observations | 120 | 40 |
| Median PFS (months) | 14.5 | 12.6 |
| Hazard ratio (95% CI)[#][*] | 0.705 (0.566, 0.877) | |
| Log-rank[*] p-value | 0.002 | |

\# Hazard ratio estimated by fitting Cox proportional hazard (PH) model to the data

\* Stratified by randomization stratification factors

Since the KM curves did not begin to separate meaningfully until approximately 12 months from randomization, there is considerable skepticism about the contribution of the combination phase of veliparib and SOC CT to the efficacy of the entire regimen, and that any observed benefit is predominantly driven by the effect of the monotherapy phase. Such concerns have been raised since clinical trials with other PARP-inhibitors (PARPi) have demonstrated excellent efficacy with only maintenance treatment. Such cross-study comparisons are however fraught with inferential issues, especially since maintenance studies of other PARPi were restricted to subjects who received and responded to CT prior to enrolment and initiating study treatment. Furthermore, substantial proportions of subjects in BROCADE3 continued their combination phase well past the point where the Kaplan-Meier estimates of PFS time began to separate and remained differentiated, and less than 40% of subjects eventually transitioned to monotherapy (see Figure 4). It is therefore incorrect to deem the combination phase unnecessary from an efficacy perspective simply based on topline results and the considerations above. In terms of safety, the SOC chemotherapy regimen is not without toxicity and combining veliparib with it further increases the treatment burden for study subjects. It is therefore important to assess the contribution of the combination phase to the overall risk-benefit of the new regimen.

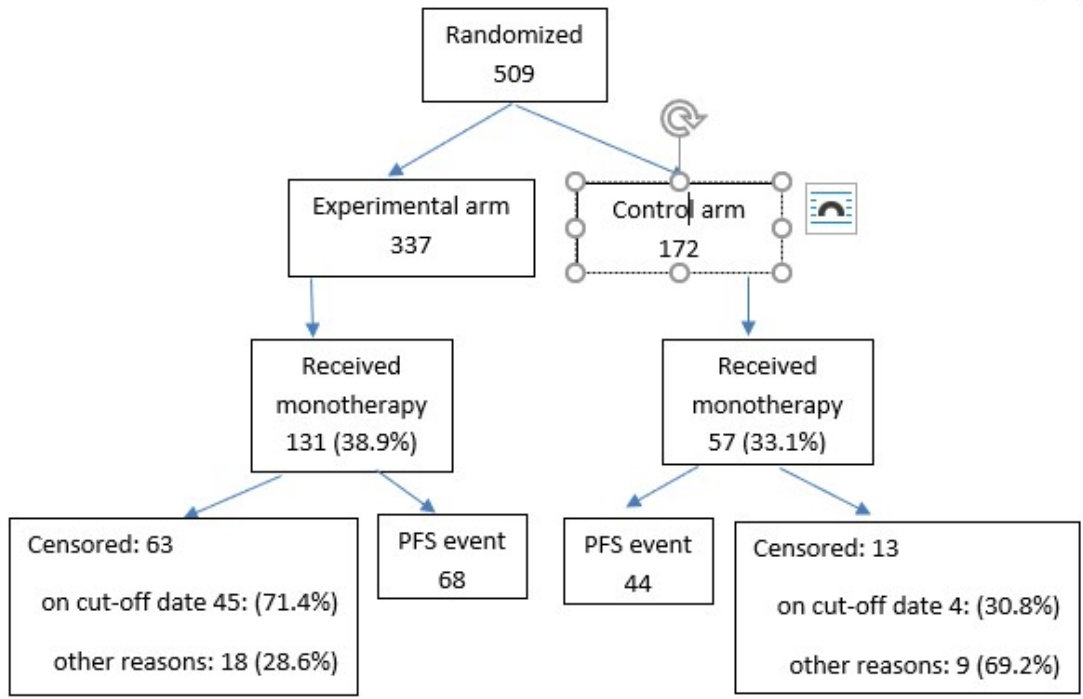

**Figure 4:  Event and censoring during monotherapy**

An important factor to note here is that the combination phase of the study was not limited to a fixed duration. Subjects could receive CT until progression or death or for as long as it remained tolerable. At the time of the primary analysis, 188 (36.9%) of the 509 subjects included in the intent-to-treat (ITT) analysis had transitioned to veliparib/placebo monotherapy comprising 57/172 (33.1%) subjects from the control arm and 131/337 (38.9%) subjects from the experimental arm. As may be expected, over time the fraction of ongoing subjects who had transitioned to monotherapy increased (see Table 2).

Table 2: Subjects transitioning to monotherapy over time

|  | Experimental Arm (N=337) | | Control Arm (N=172) | |
| --- | --- | --- | --- | --- |
| No. of Subjects at: | On treatment | On monotherapy[a] | On treatment | On monotherapy[a] |
| 1.0 year | 167 (49.6%) | 76 (45.5%) | 97 (56.4%) | 49 (50.5%) |
| 1.5 years | 119 (35.3%) | 66 (55.5%) | 54 (31.4%) | 34 (63%) |
| 2.0 years | 81 (24.0%) | 50 (61.7%) | 30 (17.4%) | 17 (56.7%) |
| 2.5 years | 56 (16.6%) | 39 (69.6%) | 22 (12.8%) | 14 (63.3%) |
| 3.0 years | 32 (9.5%) | 25 (78.1%) | 15 (8.7%) | 12 (80%) |

2.1. Percentages represent the proportion of subjects on monotherapy among those who are still on treatment in the corresponding treatment arm

## 2.2. Cox Modeling with Time-varying Covariate

To characterize the impact of transitioning to monotherapy on the overall treatment effect, Diéras et al.[2] published results of a Cox model fitted to the observed PFS times that included an indicator for initiation of veliparib/placebo monotherapy as a time-varying covariate, treatment assigned at randomization, and the interaction of these two factors (Table 3). The fitted model produced separate HR estimates describing the difference between the treatment arms during the combination and monotherapy phases, both of which favored the experimental arm, with HRs of 0.811 (95% CI: 0.622, 1.056) and 0.493 (95% CI: 0.334, 0.728) during combination therapy and monotherapy, respectively.

**Table 3:** Time-Varying Covariate Analysis of Progression-Free Survival in BROCADE 3

|  | Combination Phase | | Monotherapy Phase | |
|---|---|---|---|---|
|  | Experimental Arm | Control Arm | Experimental Arm | Control Arm |
| **PFS (Assessed by Investigator)**[a] | | | | |
| No. of subjects | 337 | 172 | 131 | 57 |
| No. of events (%) | 149 (44.2) | 88 (51.2) | 68 (51.9) | 44 (77.2) |
| Hazard ratio | 0.811 | | 0.493 | |
| (95% CI) | (0.622, 1.056) | | (0.334, 0.728) | |

C = carboplatin; CI = confidence interval; ER = estrogen receptor; No. = number; P = paclitaxel; PFS = progression-free survival; PgR = progesterone receptor

Interpretation of these estimates need to be approached with caution. For instance, the estimated HR of 0.493 during the monotherapy phase may not be interpreted as the effect of monotherapy veliparib alone compared to placebo since it is likely influenced by possible carryover or delayed effects of combination therapy or is potentially biased due to lack of balance in patient and disease characteristics among those who transition to monotherapy between the two treatment arms. Characteristics of subjects who transitioned to monotherapy in each study arm are also prognostically better than those who were originally randomized. Furthermore, the observed effect is also impacted by subjects on the two arms receiving different therapies prior to transitioning to monotherapy that can influence future outcomes differently. These considerations are germane to not only the Cox modeling but also to all other analyses of data from BROCADE3 presented in this article.

For BROCADE3, while no obvious prognostic differences were noticed between the cohorts on the two arms that transitioned to monotherapy and their proportions were reasonably similar, the abovementioned challenges to interpretation of results remain. Nevertheless, what we can say is, after adjusting for the estimated differences that emerged after onset of monotherapy, the treatment difference observed during the pre-monotherapy (combination) phase is estimated to be 0.811. For the same reasons mentioned above, this may not be interpreted as the full effect of combination treatment with veliparib.

## 3. Proposed Methodology for Tipping Point Analyses

To set up the framework for the proposed method we will adopt the common framework for analysis of a time-to-event endpoint (TTE). Considering for example (and without loss of generality) PFS in the case of BROCADE3, let $T_C$ denote the time from randomization to progression or death (PFS time) of a subject receiving the control therapy, and $T_E$ for one receiving the experimental therapy. In terms of distributions, assume $T_C \sim \boldsymbol{F}_C$ and $T_E \sim \boldsymbol{F}_E$, with corresponding survival functions denoted as $\boldsymbol{S}_C = 1 - \boldsymbol{F}_C$ and $\boldsymbol{S}_E = 1 - \boldsymbol{F}_E$. Since observations of such TTE variables are frequently taken to be right censored for the purpose of analysis in clinical trials, let us denote the right-censoring times for a subject on the control (or experimental) arm as $R_C$ (or $R_E$). The observed outcome variables used in the analysis of PFS time on the control arm can then be described as the right-censored time-to-event variable $S_C = \min(T_C, R_C)$, and the censorship indicator $\Delta_C = I(T_C \leq R_C)$, where min(.) is the minimum function and I(.) is the indicator function. For the experimental arm, we denote similar variables as $S_E = \min(T_E, R_E)$ and $\Delta_E = I(T_E \leq R_E)$.

For a well conducted clinical study, it is standard to assume that $R_C$ and $R_E$ are independent of $T_C$, $T_E$. and the treatment effect, since censorship and occurrence of events will typically be governed by independent stochastic processes, either unconditionally or conditional on a set of baseline covariates. In the parlance of missing observation this means that PFS times for censored cases may be assumed to be missing (completely) at random. When the majority of censoring occurs due to administrative reasons (e.g., data cutoff for analysis), $R_C$ and $R_E$ may be assumed to be identically distributed following a common distribution $G$ independent of $S_C$ and $S_E$.

Tipping point analysis is a common approach to assess the impact of missing observations or intercurrent events (ICE) on the analysis of a clinical study when there is reason to suspect that the missing data or ICE may have substantively influenced study findings, especially when the results are statistically significant. The approach works by imputing data that are missing or impacted by specific ICE in a progressively conservative manner (i.e., biased against the experimental arm) to reduce treatment differences until ultimately a preset threshold is crossed (e.g., statistical significance is lost). The amount of pessimism built into the imputation method is controlled by one or more parameters, and the parameter setting at which imputation-based treatment differences tip over a specified threshold (e.g., become non-significant) is then referred to as the "tipping point". (Permutt[3], Zhao et al.[4])

In causal inference, RPSFT modeling was introduced and used to address issues related to the ICE of treatment switching in the context of time to event endpoints. It provides a method for estimating and generating counter-factual times when such ICE does not occur by modifying the time to event through an adjustment factor. In the purely modeling context, this factor is estimated from the data such that the effect of the ICE is neutralized and then used to reconstruct the endpoint by removing the influence of the ICE (Robins and Tsiatis[5]), White et al.[6,7]).

The method proposed in this paper combines elements from these two techniques by first using the framework of RPSFT modeling in a novel yet natural way to account for the influence of component phases within a treatment regimen on study outcomes, and then performing a tipping-point analysis using that structure to infer about the contribution of the specific treatment phase of interest.

## 3.1 Tipping Point Analysis by Counterfactual Elicitation (TPACE)

Let us now develop and describe the mathematical framework for the method through the exercise of applying it to data from the BROCADE3 study to infer about the influence of the combination treatment phase of the new experimental regimen. To do this, let us first consider two distinct effects of the combination phase that can be of potential interest and was alluded to in Section 1.

Effect 1: The contribution of adding veliparib in the combination phase to the effect of the entire regimen; and

Effect 2: The effect of adding veliparib in the combination phase without the veliparib monotherapy maintenance phase.

*Model set-up and assessment method for Effect 1*

In the context of BROCADE3, our primary interest lies in Effect 1 and we start by describing how one would assess Effect 1 using our proposed approach. As noted before, such an effect is best estimated from a hypothetical RCT of A+B+C vs. B+C, i.e., where the full regimen (including the option to receive veliparib monotherapy) is compared to a regimen of SOC CT (without veliparib) followed by the option to receive monotherapy veliparib if disease had not progressed. The latter hypothetical control arm (B+C) was not studied in BROCADE3 and hence our first challenge is to try to simulate its outcome based on observations from BROCADE3.

To set up the mathematical framework for simulating such counterfactual data, we utilize the construct of the RPSFT model with initiation of monotherapy as the key ICE. For the control arm then, suppose that $X_C$ denotes the time from randomization until initiation of monotherapy (placebo in the actual study) or a PFS event, whichever occurs earlier. The RPSFT then postulates that the adjusted time to event (progression or death) for subjects on the hypothetical control arm can be derived as:

$$T'_C = X_C + \gamma_C(T_C - X_C) = X_C + \gamma_C Y_C.$$

Here $\gamma_C$ is the adjustment factor that modifies the time on placebo monotherapy $Y_C = (T_C - X_C)$ to what it might have been if the subject had instead received veliparib monotherapy, the counterfactual. It is reasonable to assume $\gamma_C \geq 1$ since, for subjects who received monotherapy, the time from initiation of monotherapy to a PFS event is expected be prolonged if placebo is replaced with veliparib. Also, since we take censorship to be governed by a treatment-independent stochastic process and follow an independent distribution $G$, observed or potential censoring times are expected to be unaffected when the treatment is switched.

The factor $\gamma_C$ is the critical parameter for our proposed approach to a TPA here. It controls the amount of inflation applied to the effect of placebo monotherapy on the original control arm C to mimic the counterfactual effect of the hypothetical control arm B+C. In our method, we propose to progressively increase $\gamma_C$ until a tipping point is reached. We can set this TPA threshold in a couple of ways. We can increase $\gamma_C$ until either

a) statistical significance of the PFS difference between the two arms is lost (this follows the principle of the original "tipping point" approach), or

b) the treatment difference that emerged during the monotherapy phase is "neutralized".

We will consider the value of $\gamma_C$ at which a TPA threshold is reached as representing the tipping point and describe later in this section how it can be estimated. With threshold (a) above, the reduced HR at the tipping point will be approximately equal to 0.8, the MDD. The higher the estimated tipping point of $\gamma_C$ is for this threshold, the more likely it is that statistical significance in BROCADE3 is not solely an effect of monotherapy maintenance. Similarly, for criterion (b), the HR achieved at the tipping point may be interpreted as the "minimum" benefit that can be ascribed to the combination phase after the monotherapy effect is neutralized (i.e. reduced to approximately HR = 1). Both these thresholds thereby provide an assessment of the contribution of the combination phase to the full regimen. We will elaborate on these approaches more later in the article. In this section we will describe the process for generation of the counterfactuals and the role $\gamma_C$ plays in it. Then in section 3.2, we will lay out how TPA threshold (b) may be achieved. To distinguish the tipping parameters for the two thresholds, we will denote the modifier corresponding to TPA threshold (b) as $\alpha_C$.

Reverting to the description of the actual methodology for assessing Effect 1, note that in BROCADE3, only 33.1% of the subjects on the control arm received monotherapy with placebo, i.e., for whom observed $Y_C > 0$. Thus, the adjustment factor $\gamma_C \geq 1$ may be applied directly to only these observations. Observations on the experimental arm are left unaltered. Suppose, for the $i$-th subject on the control arm who received placebo monotherapy, the observed survival time and censoring indicator were $(s_{Ci}, \delta_{Ci})$ and time to initiating monotherapy, censoring time and time from initiating monotherapy to a PFS event (unobserved when the observation is censored, i.e., $\delta_{Ci} = 0$) were $x_{Ci}, r_{Ci}$ and $y_{Ci}$, respectively. Then given that $\gamma_C \geq 1$ we find $t'_{Ci} = x_{Ci} + \gamma_C y_{Ci} \geq x_{Ci} + y_{Ci} = t_{Ci}$. Therefore, the counterfactual PFS time and censoring indicator $(s'_{Ci}, \delta'_{Ci})$ can be obtained for $\delta_{Ci} = 0$ as:

$$(s'_{Ci}, \delta'_{Ci}) = (s_{Ci}, \delta_{Ci}) = (r_{Ci}, 0)$$

since $t'_{Ci} \geq t_{Ci} \geq r_{Ci}$ and one derives

$$\delta'_{Ci} = I(t'_{Ci} \leq r_{Ci}) = I(t_{Ci} \leq r_{Ci}) = \delta_{Ci} = 0;$$

$$s'_{Ci} = \min(t'_{Ci}, r_{Ci}) = \min(t_{Ci}, r_{Ci}) = s_{Ci} = r_{Ci}.$$

In other words, the censored observations would remain unchanged. The actual PFS time had not been observed in these cases, and since a subject's PFS time is assumed to only increase if placebo is substituted with veliparib in maintenance, the simulated counterfactual observation would still be right-censored at the same censoring time and remain unchanged.

On the other hand, when $\delta_{Ci} = 1$ the actual PFS time $t_{Ci}$ would have been observed and the potential censoring time $r_{Ci}$ is unobserved. If we could impute $r_{Ci}$ or it is otherwise known, then depending on the value of $\gamma_C$ we would have:

$$\begin{aligned}(s'_{Ci}, \delta'_{Ci}) &= (t'_{Ci}, 1) & \text{if } t'_{Ci} = x_{Ci} + \gamma_C y_{Ci} \leq r_{Ci} \\ &= (r_{Ci}, 0) & \text{if } t'_{Ci} > r_{Ci}.\end{aligned} \quad (1)$$

Thus, given a putative value of the effect modifier $\gamma_C$, to impute the counterfactual PFS observation $(s'_{Ci}, \delta'_{Ci})$ for a subject on the control arm who had experienced a PFS event after initiating placebo monotherapy, one needs to simulate the unobserved censoring time $r_{Ci}$ conditional on the observation that $t_{Ci} \leq r_{Ci}$. The conditioning ensures that, when $\gamma_C = 1$ (i.e., the original observations are left unaltered), the hazard ratio between treatment arms would also remain unchanged. There are then essentially two ways to simulate from this conditional distribution:

1. For a well-conducted study where a majority of censoring occurs due to data cutoff for analysis with low loss to follow-up rates, $r_{Ci}$ may simply be imputed using the time from randomization to the data cutoff date where necessary. This in effect assumes that the counterfactual disease progression or death of these subject would have been observed were it to occur by the data cutoff date.

2. If censoring due to reasons other than administrative data cutoff are non-negligible, then it is better to obtain an estimate of the common, independent censorship distribution $G$ by fitting an appropriate parametric or semiparametric survival model after reversing the censorship indicator. This reduces simulation of the censoring times to the simple task of sampling from the fitted model using rejection sampling (to simulate an observation $r_{Ci} \geq t_{Ci}$).

Once the unobserved censoring times $r_{Ci}$ have been simulated, for any given value of $\gamma_C$ the counterfactual observations for PFS can be generated using equation (1).

*Model set-up and assessment method for Effect 2*

If interest lies in the pure effect of the combination phase, then the best way to evaluate it is through a RCT comparing veliparib plus SOC CT (without an option to receive veliparib monotherapy) to SOC CT only. Thus if interest lies in Effect 2, then the control arm in BROCADE3 could be left as is, but observations on the treatment arm would need to be adjusted to simulate what they would be if the option to receive monotherapy was absent. In this case, we will set up the RPSFT model as follows:

$$T'_E = X_E + \gamma_E Y_E.$$

Since veliparib monotherapy is believed to be effective, in a reciprocal formulation we will now assume $0 < \gamma_E \leq 1$. In other words, the time from initiation of monotherapy to a PFS event would be shortened if veliparib is replaced with placebo in the monotherapy phase. Censoring times (observed or unobserved) however, will once again remain unaffected due to the independence of $G$.

To estimate the tipping points for Effect 2 using the same TPA thresholds as with Effect 1, the factor $\gamma_E$ can now be progressively decreased until the corresponding threshold is reached. The lower the estimated tipping points of $\gamma_E$ is, the more likely it is that the combination phase of veliparib plus SOC CT has a substantive effect by itself.

Modifications using the adjustment factor $\gamma_E$ in this case will apply to the roughly 40% of the subjects on the veliparib arm in BROCADE3 who received monotherapy, and observations on the control arm will remain unaltered. Adopting similar notation as before, given that $0 \leq \gamma_E \leq 1$ for Effect 2, we find the counterfactual time to event will be shortened since $t'_{Ei} = x_{Ei} + \gamma_E y_{Ei} \leq x_{Ei} + y_{Ei} = t_{Ei}$ and observed events will remain as events with shortened duration. Thus, the adjusted survival time and censoring indicator $(s'_{Ei}, \delta'_{Ei})$ can be obtained for $\delta_{Ei} = 1$ as $(s'_{Ei}, \delta'_{Ei}) = (x_{Ei} + \gamma_E y_{Ei}, 1)$ since $t'_{Ei} \leq t_{Ei} \leq r_{Ei}$.

On the other hand, for $\delta_{Ei} = 0$, it is uncertain if the originally censored PFS time $t_{Ei}$ would have been observed if hypothetically veliparib was replaced by placebo as monotherapy and PFS time was shorter. If we can use an estimate of $F_E$ conditional on the observation that $t_{Ei} > r_{Ei}$ to impute $t_{Ei}$, then depending on the value of $\gamma_{Ei}$ we would have:

$$(s'_{Ei}, \delta'_{Ei}) = (r_{Ei}, 0) \quad \text{if } \gamma_E(t_{Ei} - x_{Ei}) = \gamma_E y_{Ei} > r_{Ei} - x_{Ei};$$

$$= (t'_{Ei}, 1) \qquad \text{if } t'_{Ei} \leq r_{Ei}. \qquad (2)$$

Thus, for Effect 2, one needs to generate the counterfactual PFS observation $(s'_{Ei}, \delta'_{Ei})$ for each subject on the experimental arm who had initiated veliparib monotherapy but were censored for the primary analysis of BROCADE3. The unobserved event times $t_{Ei}$ in these cases need to be simulated conditional on $t_{Ei} > r_{Ei}$. However, the time to event (progression or death) from end of combination therapy, $Y_{Ei} = T_{Ei} - X_{Ei}$, for such subjects cannot be approximated by a degenerate (point-mass) distribution as in Approach 1 for Effect 1. Our preferred approach in this case is to use a fitted survival model to impute $y_{Ei}$ similar to Approach 2 for Effect 1 conditional on $y_{Ei} > r_{Ei} - x_{Ei}$ using a rejection sampling scheme.

In the following sections we will refer to the methods proposed in this section as the *Tipping Point by Counterfactual Elicitation (TPACE)*.

## 3.2 Relationship between RPSFT Modeling, Tipping Points and Cox Regression

We will now algebraically describe the relationship between the RPSFT modeling structure and and HRs within Cox modeling that forms the basis for our TPA with respect to achieving threshold (b) stated above. It also allows us to comparatively assess the results of the various analyses and gain better insights.

Suppose $\widehat{S}_E(t)$ and $\widehat{S}_C(t)$ denote the estimated survival functions for the experimental and control arm at time $t$. Then, based on the HRs estimated through Cox proportional hazards model fitted on all data of BROCADE 3, we obtain $\widehat{S}_E(t) = [\widehat{S}_C(t)]^{0.705}$. Similarly, based on the results of the Cox model with time-varying covariate for BROCADE3 presented above, we have $\widehat{S}_E^A(u) = [\widehat{S}_C^A(u)]^{0.811}$, where $\widehat{S}_E^A$ and $\widehat{S}_C^A$ are the estimated survival functions for the experimental and control arm for the duration of the combination phase (i.e., time from randomization to onset of monotherapy or occurrence of a PFS event). Similarly, one would obtain $\widehat{S}_E^B(v) = \left[\widehat{S}_C^B(v)\right]^{0.493}$ where $\widehat{S}_E^B$ and $\widehat{S}_C^B$ are estimated survival functions for the experimental and control arm for time from onset of monotherapy to a PFS event.

Suppose $\alpha > 1$ is a time-scaling factor that neutralizes the difference in the PFS distributions during the monotherapy phase estimated via the Cox model, such that:

$$\widehat{S}_E^B(v) = \left[\widehat{S}_C^B(v)\right]^{0.493} \cong \widehat{S}_C^B(v/\alpha).$$

We can then write the relationship between survival estimates $\widehat{S}_E$ and $\widehat{S}_C$ as:

$$\widehat{S}_E(t) \cong \left[\widehat{S}_C(x_C + \alpha(t - x_C))\right]^\theta,$$

where $\theta$ represents the HR that remains after removing the difference between treatment and control from the monotherapy phase via a RPSFT model-based construct. For example, setting $\alpha = 2$ in the relationship between $\theta$, $\widehat{S}_E$ and $\widehat{S}_C$ above yields $\theta \cong 0.8$, indicating that doubling the survival time during monotherapy on control arm would reduce the treatment effect (hazard ratio) between the two treatment arms from 0.705 to roughly 0.8.

In terms of the implementation of TPA threshold (b) for the assessment of Effect 1, the equations above simply entails identifying a tipping value of the inflation factor $\alpha$ for PFS times during the placebo monotherapy phase of the control arm until the estimated monotherapy HR of 0.493 is reduced to 1. We will then simulate counterfactuals using this tipping value of the inflation factor as described earlier and estimate the reduced overall HR at the tipping point, which may be viewed as the minimum difference that emerges during the combination phase.

Similarly, in the context of Effect 2, one would identify a tipping value of a shrinkage factor for PFS times during the veliparib monotherapy phase on the experimental arm that reduces the estimated monotherapy HR to 1. We can then simulate counterfactuals using this tipping value of the shrinkage factor and estimate the reduced overall HR. The equations corresponding to Effect 2 can be easily found in a way similar to those for Effect 1 (and are hence not provided here).

## 4. Application of the Tipping Point Analyses

We will now illustrate the methods described in Section 2 using the data from the BROCADE3.

*TPA for assessment of Effect 1 based on counterfactual PFS times on the control arm*

Tipping point analysis using the method proposed earlier in this paper was performed including data from all ITT subjects (N=509). An inflation factor $\gamma_C$ ($\gamma_C > 1$) was applied only to those who transitioned to monotherapy on the control arm. Since overall, the experimental regimen is more efficacious, a lower percentage of events was observed on the experimental arm compared to the control arm during the monotherapy phase, as depicted in the study consort diagram (Figure 3). This results in a larger proportion of censored observations on the experimental arm compared to control during this phase. Among all censored observations across both treatment arms, PFS times for a majority of subjects (64.5%) were censored on the cut-off date. With this in mind, we simulate the counterfactual observations to assess Effect 1 by simply imputing the time from randomization to the cut-off date as the unobserved censoring time for any subject who experienced a PFS event following monotherapy in BROCADE3. However, if the proportion of censorship prior to the cut-off date is relatively high, separate estimation of the censoring distribution would be necessary. Starting from $\gamma_C = 1$, where results are same as the primary analysis, we performed a grid-search by progressively increasing the value of $\gamma_C$ to find TPA threshold (a) for PFS (Figure 5). Similarly, the value of $\alpha_C$ can be modulated to identify TPA threshold (b) for PFS (Figure 5).

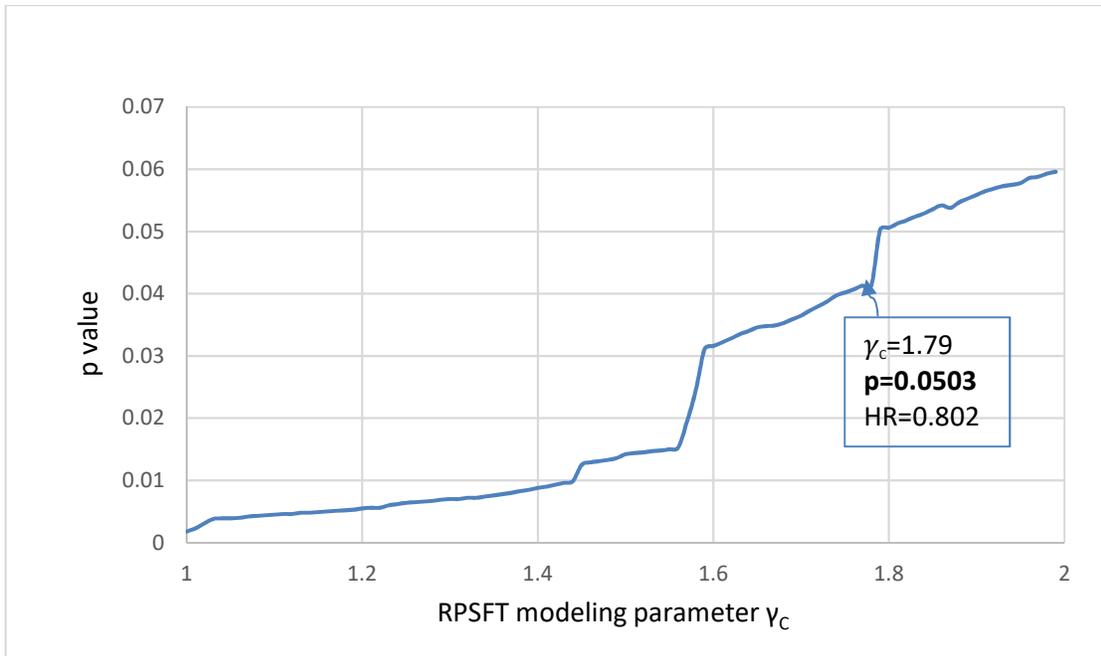

**5A.** Plot of counterfactual p-values vs. values of $\gamma_C$ for the control arm

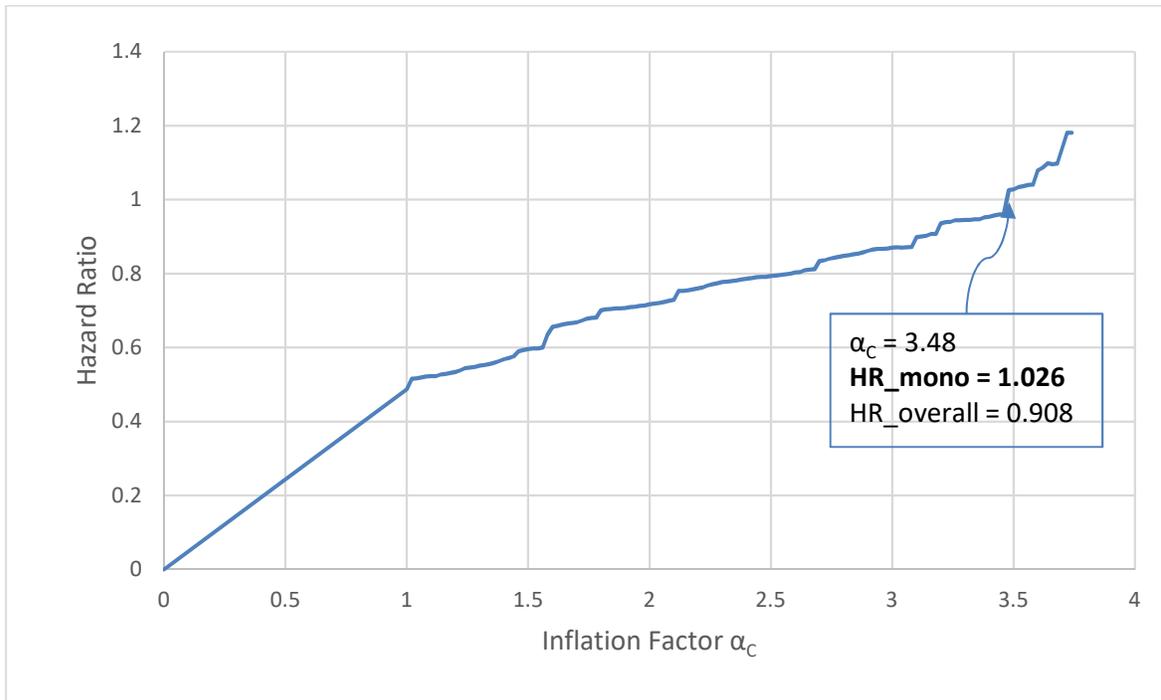

**5B.** Plot of counterfactual Hazard Ratios during monotherapy vs. values of $\alpha_C$ for the control arm

**Figure 5.** **Results of tipping point analysis for Effect 1 (A) up to the HR of overall treatment effect reduces to MDD and (B) up to neutralization of the monotherapy effect**

*TPA for assessment of Effect 2 based on counterfactual PFS times on the experimental arm*

Tipping point analysis for assessing Effect 2 was performed with the shrinkage factor $\gamma_E$ ( $0 < \gamma_E < 1$) applied only to those who transitioned to monotherapy on the experimental arm. Counterfactual PFS times for subjects who were randomized to the experimental arm and went on to receive monotherapy needed to be simulated conditional on observed data in this case. To illustrate the approach, we assumed and fit an exponential distribution for the subjects time from initiation of monotherapy to progression or death. We note that by virtue of the lack-of-memory property of the exponential distribution, for subjects censored during the monotherapy phase, the unobserved time from censoring to potential progression or death would also follow the same distribution. For these subjects, we therefore simulated the unobserved time from censorship to event using the fitted exponential distribution and added it to the observed $s_{Ei}$ to obtain $t_{Ei}$. With the time to event, $t_{Ei}$ values, thus obtained for all subjects on the experimental arm, we shrink them by the factor $\gamma_E$ following the rules described in preceding section. For subjects who were originally censored during monotherapy if the shrunk, imputed time to event is smaller than the observed censoring time, then the counterfactual observation for that subject will be an imputed event with PFS time set equal to the imputed time to event. Once again, starting from $\gamma_E = 1$ and $\alpha_E = 1$ (where results are same as that of the primary analysis), grid-searches by decreasing the value of $\gamma_E$ and $\alpha_E$ were performed to find TPA thresholds (a) and (b) for PFS, respectively (Figure 6).

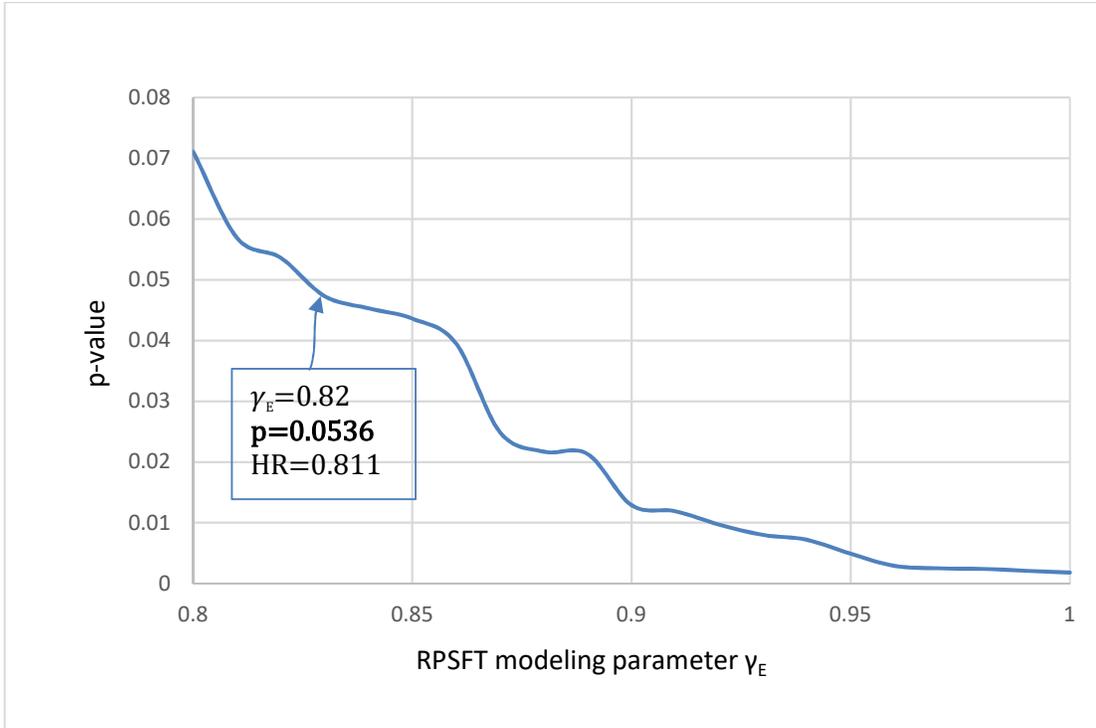

**6A.** Plot of counterfactual p-values vs. values of $\gamma_E$ for the experimental arm

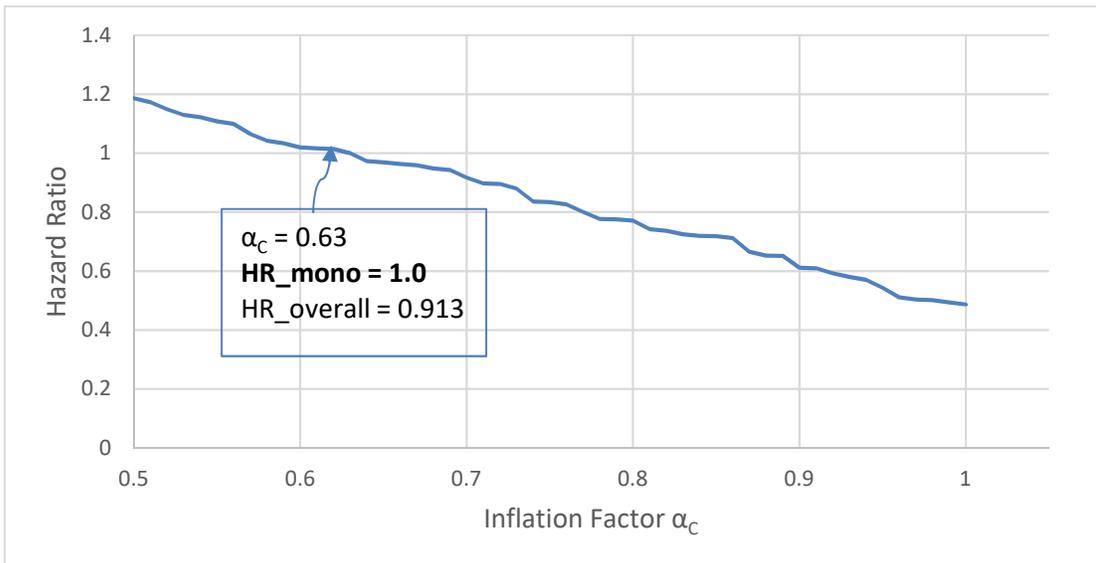

**6B.** Plot of counterfactual Hazard Ratios during monotherapy vs. values of $\alpha_C$ for the experimental arm

**Figure 6.** **Results of tipping point analysis for Effect 2 (A) up to the HR of overall treatment effect reduces to MDD and (B) up to neutralization of the monotherapy effect**

*A naïve variant to "proper counterfactuals" based TPA*

For time to event endpoints, under standard assumptions statistical information is proportional to the number of observed events. In the tipping point analysis with proper counterfactuals proposed in Section 2 and as illustrated with the BROCADE3 data, depending on the value of $\gamma$ the number of events in the simulated counterfactual can change from what was actually observed, thereby changing the information content of the observed trial data. As such, we perform a naïve sensitivity analysis where we leave the information content of the dataset unaltered by keeping the number of events unchanged for various values of $\gamma$.

In this analysis, the factor for detecting the tipping point was multiplied to the survival times of all subjects who had monotherapy, regardless of whether they had an event or were censored. Algebraically speaking, we apply the inflation factor $\gamma_C$, ($\gamma_C > 1$) to all subjects on control arm who received monotherapy to obtain $\tilde{s}_{Ci} = x_{Ci} + \gamma_C(s_{Ci} - x_{Ci})$ regardless of whether $\delta_{Ci} = 1 \text{ or } 0$, implying $\tilde{s}_{Ci} > s_{Ci}$ for all subjects on the control arm. As a consequence, time to censoring also changes with the choice of $\gamma_C$, and the censoring distribution may now become somewhat dependent on the treatment effect. This data was then augmented with data from subjects who never received monotherapy, and a survival analysis was performed comparing the new (shrunk or inflated) PFS times between the treatment arms to provide sensitivity to the assessment of Effect 1 described above. Results are presented under "Naïve Variant" in Table 4.

Alternatively, we can apply a shrinkage factor $\gamma_E$, ($0 < \gamma_E < 1$) to all subjects on experimental arm with a monotherapy phase to obtain $\tilde{s}_{Ei} = x_{Ei} + \gamma_E(s_{Ei} - x_{Ei})$ regardless of whether they had an event or not, i.e., $\delta_{Ei} = 1 \; or \; 0$. This implies $\tilde{s}_{Ei} < s_{Ei}$ for all subjects on the experimental arm and provides a sensitivity to the assessment of Effect 2 (Table 4).

Table 4: **Results of tipping point analysis of BROCADE3 evaluating effect of the combination phase**

| Effect / Method of estimation | Adjustment factor γ at tipping point[a] | Average No. of Events | Hazard ratio at tipping point | P-value at tipping point[a] |
|---|---|---|---|---|
| Assessment for Effect 1 | | Tipping parameter $\gamma_C$ | | |
| Proper counterfactuals | 1.79 | 342 | 0.802[a] | 0.0503 |
| Naïve Variant | 1.95 | 349 | 0.806[a] | 0.0526 |
| | | Tipping parameter $\alpha_C$ | | |
| TPACE Method | 3.48 | 335 | 0.908 | |
| Assessment of Effect 2 | | Tipping parameter $\gamma_E$ | | |
| Proper counterfactuals | 0.82 | 379 | 0.811[a] | 0.0536 |
| Naïve Variant | 0.51 | 349 | 0.805[a] | 0.0501 |
| | | Tipping parameter $\alpha_E$ | | |
| TPACE Method | 0.63 | 397 | 0.913 | |

a. In all cases, tipping point is determined at value of the modeling parameter γ (either adjustment factor $\gamma_C$ or $\gamma_E$) with minimum p > 0.05 (2-sided).

One may be intrigued by the observations that:

i. the estimated HR for the monotherapy therapy phase from the Cox model (in Section 2.1) is 0.493 is rather close to the adjustment factor value of 0.52 for the monotherapy phase of the treatment arm at the tipping point obtained while assessing Effect 2 in sensitivity analysis 1 (or the reciprocal of the factor value 1.94 for the mono phase of the control arm obtained at the tipping point for assessing Effect 1).

ii. This value 0.8 of the reduced HR that arises out of the equations in Section 3.2 when α = 2 is similar to the MDD of the study as well as the HR estimate of 0.811 for the monotherapy phase from the Cox model in Section 2.1.

Such similarities are, however, simply coincidental in the context of BROCADE3 as far as we can tell.

## 5. Discussion

As described earlier, we view our motivating example of BROCADE3 as a study of A+B+C vs. C where C represents the SOC CT, A is veliparib used as part of a combination phase with CT and B is veliparib monotherapy at a higher dose given after CT is discontinued (Figure 1). On the control arm both low- and high-dose veliparib is replaced by placebo. Our proposed tipping point approaches are an attempt to gain insights about the utility of A as a part of the full experimental regimen in the treatment of patients with BRCA+ HER2- metastatic breast cancer. Given the study design, neither the effect of A nor B can be fully isolated since the effect of B will always remain susceptible to potential carryover effects of A. To do so, one would need separate RCTs including study arms A+C or B+C that were not included in the study.

Our method is not a substitute for such RCTs but tries to provide an assessment of the utility of A using the temporal separation of the two phases. This is done through some tipping point analyses of simulating data from the missing, hypothetical study arms conditional on the observed study data in a statistically sound manner using an RPSFT modeling structure. In Section 2, we described two effects (estimands) to formalize our goal. Effects 1 and 2 corresponds to the comparative effects of A+B+C vs B+C (the contribution of A) and of A+C vs C (the effect of adding A to C), respectively.

From the perspective of clinical practice, Effect 1 seems far more important to assess than Effect 2 since only the full regimen (not A+C) is being considered for treatment of patients, but both can be useful to understand the utility of A in some ways. For example, if Effect 2 is assessed not to be clinically meaningful then one could infer that >60% of the study population (percentage of those who did not receive B in BROCADE3) would not benefit from receiving veliparib.

A Phase 2 study of veliparib only as a combination to SOC CT (without an option to be used as monotherapy), i.e. a comparison of A+C vs. C, was conducted in a similar patient population and yielded an estimated HR of 0.789 (Han et al.[8]). Findings of the Cox regression discussed in Section 2.1 show that after adjusting for the estimated treatment differences that emerged during the monotherapy phase, the difference that emerged during the combination phase was estimated to be HR=0.811 and is consistent with the Phase 2 findings. Assessment of Effects 2 via our method using counterfactuals indicate that, when influenced by prior A, the outcomes following initiation of B must be about 18% ($\gamma_E$ =0.82) shorter for statistical significance to be lost (i.e., the overall HR reduces close to the MDD of 0.8). Assessment of Effect 1 indicates that without the influence of prior treatment with A, outcomes following onset of monotherapy placebo must be inflated by 79% ($\gamma_C$ =1.79) to reach the same TPA threshold. Given these findings, it seems reasonable to assess that the 20% reduction in hazard that was observed in Phase 2 for A+C is not negligible and the statistical significance of the full regimen (A+B+C) is unlikely to be solely driven by B.

In terms of TPA threshold (b) in the context of assessing Effect 1, PFS time from onset of placebo monotherapy needs to be inflated by nearly 250% for the HR for the monotherapy phase to be neutralized in the absence of prior treatment with A, and under such inflation the residual HR is roughly 0.9. On the other hand, in the presence of prior treatment with A (in the context of Effect 2) the PFS times following onset of B on the experimental arm needs to be shrunk by approximately 37% to neutralize the effect emerging during the monotherapy phase, including any carryover effect of the combination phase, while also reducing the overall HR to 0.9. The 10% reduction in hazard that remains after neutralizing the HR for the monotherapy phase in both these cases may be seen as the effect that can be ascribed to A without accounting for any carryover or synergistic effects it may have. While the interaction of the effects of A and B may be positive, negative or neutral, the positive trend makes such effects more likely to be positive or synergistic.

The naïve variant to our proposed TPA in Section 4 is included to satisfy the curiosity of readers who might wonder about the impact of altering the total information content (i.e., total events) and power in the simulated counterfactual study data. While this variant is simple to implement, it invalidates the independence of the treatment effect and the censoring distribution $G$. Other approaches that may leave the total number of events fixed (such as leaving the censored observations unchanged) are also problematic since they alter the underlying time-to-event survival functions. Hence or otherwise, these approaches do not provide any meaningful assessment of the estimands of interest in our opinion and are not recommended.

It is important to note now that the simulation of counterfactuals (replacing placebo monotherapy with veliparib monotherapy or vice versa) in our proposed methods should be viewed as imputing effects that are conditional on the rest of the treatment regimen received in both phases by the subject remaining the same. To interpret these as isolated, unconditional effects one would need to make strong simplifying assumptions such as:

a. Any carryover effect of A is washed out prior to initiation of B and is the same irrespective of the duration of A or the actual treatment received,

b. Any effect of B begins from the day it is initiated and is the same irrespective of its duration or actual treatment (e.g., dose) received, and

c. There is little or no interaction between of the effects of A and B.

As with the case of BROCADE3, it is usually not possible to determine the validity of such assumptions in any conclusive manner. For example, the time to wash out of the carryover effect of A is usually intractable and cannot be reliably estimated from study data alone. This is not only true for our proposed method, but also for the conventional statistical methods (such as those presented in Section 2.1).

The illustration of our method in this paper has focused on assessing the effect of A. It is natural to then ask if it is possible to assess the contribution of B to the full regimen in a similar way, should it be the focus of interest. The answer to this depends on the design of the study. For BROCADE3, this is not feasible since ending combination treatment and transitioning to monotherapy is dependent mainly on the tolerability of the combination phase rather than any efficacy outcome. Thus, there is no meaningful information to generate counterfactual observations for a hypothetical A+C treatment arm for comparison with A+B+C. Had transition to monotherapy been premised on an efficacy outcome related to PFS following combination A+C, a modeling structure could be formulated for exploratory assessment of the contribution of monotherapy veliparib. A hypothetical B+C vs C comparison may be formulated.

In the same vein, we would emphasize that, despite the challenges in isolating the effects of interest (and related estimands) from the A+B+C vs C study design, certain design considerations can help in simplifying the interpretation of the findings of our TPA and other model-based analyses. One such consideration would be to have well-defined specifications for when the maintenance phase would start, e.g. meeting a sustained response criterion, receiving a maximum number of combo cycles, or experiencing some unacceptable safety or tolerability issue. One might also explore the possibility of randomization at the beginning of the maintenance phase for a better design, although this adds more complexities to analysis.

In the context of treatment crossover, the Accelerated Failure Time (AFT) model is sometimes used in the same way as the RPSFT model for time to event endpoints to adjust for treatment switching (Latimer NR, et al.[9,10]). It is easy to see that one can adapt the structure of the AFT model in the same way as we have used the RPSFT modeling structure to achieve the similar inferential goals. Our recommendation then is that, whichever method is chosen (adaptation of the RPSFT or the AFT structure), one should employ it consistently for assessment of both of the tipping points for inferential purposes.

Finally, we conclude with a few technical notes about our proposed methodology. First, we emphasize the importance of imputing time-to-event observations conditional on the observed data (i.e., sampling from conditional distributions). This is critical to ensure that, when the value of the RPSFT model parameter $\gamma$ is set to 1, i.e. the observed dataset is left unchanged, the effect estimates will match the actual findings of the study. Second, for Effect 2, our approach is dependent on estimating and sampling data from the time to event distribution for subjects with censored observations. We advise paying close attention to the estimation and simulation steps during implementation of our method. If imputation is needed for only a small percentage of subjects (e.g., only a few observations are originally censored), we would recommend using multiple imputation to obtain a robust estimate of the tipping point (and its variability) since there can be considerable sampling variability in single rounds of simulated data leading to greater uncertainty in estimating the tipping point. Also, when the time to event distribution is influenced by nonproportional hazards (e.g., late separation) and common parametric models do not provide a good fit to the observed data, a bootstrapping approach may be simpler and preferable. We also draw attention to the non-proportionality of hazards observable in Figure 3 where the separation of hazards is low until 12 months, most appreciable between 12-24 months, and less so after that. For a proportional hazards model that only includes treatments, with perhaps stratification factors as strata, the estimated hazard ratio can be reasonably viewed as an average hazard ratio regardless of departures from proportional hazards. Thus, we point that the methodology in the manuscript and its application to the motivating example are still of interest regardless of any suggestion of departure from proportional hazards. This may however be a topic of further research.

**Acknowledgments:** The authors are deeply thankful to Professor Gary Koch, David Maag (MD), Claire Tsao (PhD) and other colleagues from the Statistics and Clinical departments at AbbVie Inc. for their thoughtful inputs that this research work has greatly benefitted from.